# COMPARISON OF BRAIN CONNECTOMES USING GEODESIC DISTANCE ON MANIFOLD: A TWINS STUDY

*A. Yamin[1,2]\*, M. Dayan[1], L. Squarcina[3], P. Brambilla[4], V. Murino[1,5], V. Diwadkar[6]✆, and D. Sona[1,7]✆\*.*

[1]Pattern Analysis and Computer Vision, Istituto Italiano di Tecnologia, Genova, Italy.
[2]Department of Electrical, Electronics and Telecommunication Engineering and Naval Architecture Università degli Studi di Genova, Italy. [3]Scientific Institute IRCCS "E. Medea", Bosisio Parini, Italy.
[4]Fondazione IRCCS Ca' Granda Ospedale Maggiore Policlinico, Università di Milano, Italy.
[5]Department of Computer Science, Università di Verona, Italy.
[6]Department of Psychiatry & Behavioral Neuroscience, Wayne State University, Detroit, MI, USA.
[7]Neuroinformatics Laboratory, Fondazione Bruno Kessler, Trento, Italy
✆Equal contribution of authors. \* muhammad.yamin@iit.it, diego.sona@iit.it

## ABSTRACT

fMRI is a unique non-invasive approach for understanding the functional organization of the human brain, and task-based fMRI promotes identification of functionally relevant brain regions associated with a given task. Here, we use fMRI (using the Poffenberger Paradigm) data collected in mono- and dizygotic twin pairs to propose a novel approach for assessing similarity in functional networks. In particular, we compared network similarity between pairs of twins in task-relevant and task-orthogonal networks. The proposed method measures the similarity between functional networks using a geodesic distance between graph Laplacians. With method we show that networks are more similar in monozygotic twins compared to dizygotic twins. Furthermore, the similarity in monozygotic twins is higher for task-relevant, than task-orthogonal networks.

## 1. INTRODUCTION

Connectomics is a new research area in neuroimaging, enabling measurement and investigation of the associations between different regions in the human brain [1]. In connectomics, the brain is represented by a graph, and indexes from graph theory are commonly used to examine connectivity in brain networks [2]. *Functional connectivity* (FC) is usually built computing Pearson correlation between average time-series (BOLD activity) of brain regions defined by a suitable atlas. Then, metrics are defined to analyze the overall functional brain network, typically investigating differences between groups of subjects. How to decide on an appropriate choice of metric has been an active research area in recent years, though, almost all solutions rely on graph theory indexes (e.g., characteristic path length, modularity, etc.) [3]. These indices provide an overall description of the network, but do not generally measure the difference between networks.

The simplest approach to cope with this issue is to directly compare the FC matrices using an Euclidean metric. However, the Euclidean space may not fully describe the actual geometry of the data, suggesting that the Euclidean distance may not be the most appropriate metric to use. In response to this issue, recent approaches promote the use of geodesic distances defined on a network space, allowing to take into account the complex geometric nature of graphs, which can be considered as points in a complex manifold. In particular, a transformation of graph representation into a space of *symmetric, positive definite* (SPD) matrices has been recently suggested due to its geometrical properties (formation of Riemannian manifold) [4,5], and the possibility to adopt geodesic distances defined on it.

In this paper we specifically discuss an approach to analyze task-based fMRI data allowing the comparison of functional brain networks between *monozygotic* (MZ) and *dizygotic* (DZ) twins. To accomplish this, we resort to graph Laplacian, making the matrices semi-positive definite. This allowed us to compare the connectivity in terms of a geodesic distance, defined on the manifold of graphs. We computed different metrics to characterize the differences between graphs, including a component of Frechet distance [6], a geodesic distance on semi-positive definite matrices, and the Euclidean distance as baseline for comparison.

All our analyses have been performed on a task-based fMRI dataset acquired in twins, thus constituting a study of the relationship between genetic heritability and functional brain networks. The aim of this project is therefore to assess whether there is any effect of a specific task on brain network of two groups of twins, and analyze the influence of genetics on the functional connectivity. Our study follows a limited series of studies using fMRI in twins. In [7], fMRI and behavioral assessment were used to observe the brain structure and functioning in 26 MZ twin pairs discordant for handedness. These observations showed important correlations between language-specific functional laterality in inferior and middle frontal gyri, and anterior corpus callosum. In [8], lateralization for language, spatial judgment and face processing was assessed in 42 pairs of MZ twins (21 discordant and 21 concordant pairs for handedness) using fMRI, showing a stronger genetic influence on language asymmetry in concordant twins. In [9], 104 Pairs of twins (47 MZ, 57 DZ) fMRI were observed to examine the neural correlation in children but no genetic effect was found.

In our experiment, fMRI data were collected with the Poffenberger paradigm, traditionally used to measure interhemispheric transfer time (IHTT) [10] and more recently used to investigate fMRI responses.

## 2. MATERIAL AND METHODS:

### 2.1. Data set:

Thirteen twin pairs (7 MZ, 6 DZ,) were recruited from the population-based Italian Twin Registry. fMRI data were acquired on a 3-Tesla MR imaging unit Siemens Allegra system (Siemens, Erlangen, Germany) with a standard head coil. T2*-weighted images were acquired using a gradient-echo EPI-BOLD pulse sequence (TR: 2000 ms; TE: 30 ms; flip angle 75°; FOV: 92x192; 31 axial slices; thickness: 3 mm; in-plane: 3 mm2; matrix: 64x64). High-resolution MPRAGE T1-weighted structural images were acquired in the same session (TR: 2300 ms; TE: 3.93 ms; flip angle 12°; FOV: 256x256; 160 axial slices; slice thickness: 1 mm; matrix 256x256). fMRI scans with right and left hand were acquired consecutively in two separate scans.

### 2.2. Pre-Processing:

fMRI pre-processing was performed using a combination of shell and MATLAB scripts. The structural images (skull stripped) were processed using the FreeSurfer (http://surfer.nmr.mgh.harvard.edu/) recon-all command-line tool for parcellating the brain to create *grey matter* (GM), *white matter* (WM) and *cerebro-spinal fluid* (CSF) tissue masks. fMRI images were processed in MATLAB SPM 12. Processing includes motion correction followed by co-registration of the EPI images with an anatomical reference, applying a deformation model to normalize EPI images with an MNI template, and smoothing of images with a 4mm FWHM Gaussian filter. The $T_1$ image of each subject was nonlinearly transformed to the $T_1$ MNI152 template with FSL FNIRT [11] to match the GM mask of each subject with template. Then the GM mask of each subject was matched with the Automated Anatomical Labeling atlas [12] (90 ROIs in the Cerebrum) and applied to the processed fMRI to extract the time-series signal using FSLMEANTS. These 90 ROIs were clustered in two groups based on prior fMRI studies [13]. One group included task-relevant visuomotor (MV) ROIs (28 regions), while the other sub-group included the complementary task-orthogonal non-visuomotor (NMV) ROIs (62 regions).

The 90x90 functional connectivity matrices *W* were computed using Pearson correlations between time series, computing the r-z transform to normally distribute the data, and only retaining the positive correlations only, as commonly performed in FC analysis [14].

### 2.3. Graph Laplacian & Riemannian Manifold:

Given a symmetric undirected weighted graph *W*, it is possible to define a graph Laplacian *L*, which enjoys some properties useful for our problem. Specifically, we used the Normalized Symmetric Laplacian [15] defined as:
$$L = D^{-1/2}(D-W)D^{-1/2} \qquad (1)$$
Where $D=\text{diag}(\sum_j w_{ij})$ is the degree matrix of *W*. Graph Laplacian matrices are always symmetric and positive semi-definite, and with a small regularization they become SPD. Thus, they form a Riemannian manifold, enabling the analyses of FC matrices on a manifold instead of a vector space [16]. A distance metric based on Euclidean distance is suboptimal when applied to SPD matrices [17], because as mentioned, this metric is not responsive to the geometry in the data. To take full advantage of the manifold structure of SPD matrices, it is essential to consider some geodesic distance which measures the shortest path between two points (in our case, two matrices, one for each individual in the twin pair) along the curve of the manifold. The conceptual difference between Euclidean and geodesic distance is represented in Figure 1.

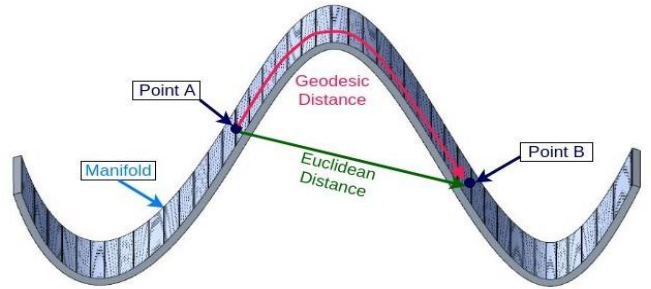

**Figure 1.** The difference between Euclidean distance of two points (green straight line) and the corresponding geodesic distance (red curve along the manifold).

### 2.4. Frechet Distance:

Many geodesic distances are defined on the Riemannian manifold of SPD matrices. For example, Log-Euclidean distance [18] is based on the difference between log values of two points. Similarly, the Stein divergence is a good approximation even if not strictly a Riemannian metric. However, these metrics can be applied only to SPD matrices, hence, Laplacian matrices need to be regularized, a process that introduces a small bias in the data, that might impact the final result.

We used, instead, a component of the Frechet distance. As in [6] the Frechet distance is defined between two Gaussian distributions, which is built on two components. One component considers the means, and the other component is between the covariance's defined as:

$$d_f^2(C_x, C_y) = tr[C_x + C_y - 2(C_x C_y)^{1/2}] \qquad (2)$$

Where $C_x$ and $C_y$ are the covariance's of the two distributions and $tr$ is the trace operator. It has been proven that the distance expressed by eq. (2) is a metric on covariance, hence, on positive semi-definite matrices. The advantage of the Frechet metric is that it allows computation of the geodesic distance between two graph Laplacians without introducing the regularization.

### 2.5. Statistical Analysis:

Frechet metric was computed for each pair of twins (MZ and DZ) providing group-wise statistics to investigate the impact of genetics on functional connectivity of task-relevant and/or task-orthogonal networks. Our working hypotheses was that connectivity between MZ twin pairs would be more similar than between DZ pairs, and particularly for the MV sub-network. In addition to the Frechet distance, we also computed Euclidean distance to highlight the sensitivity and validity of our approach.

Both Frechet and Euclidean distances are affected by the size of the matrices (i.e., the number of nodes in the graphs). Specifically, we observed that they are affected by factors $N^2$ and $N$ respectively ($N$ is the number of regions). Hence, distances were normalized to ensure analyses not confounded by network size.

A statistical analysis was then performed on distances to validate the working hypothesis. In particular, we used the Wilcoxon rank sum non-parametric test to avoid assumptions of normality on data distribution given the relatively modest sample size.

### 3. RESULTS

Figure 2 shows the comparison of Euclidean and Frechet distances. As seen the Frechet distance provides greater sensitivity for differentiating MZ from DZ based on similarity *between the task-relevant MV network*. Further analyses were therefore restricted to the Frechet distance. Analyses were undertaken accounting for Zygocity (MZ vs DZ), the response hand used in the task (Right Hand vs Left Hand) and ROI (MV vs NMV)

The plots in Figure 3 are separated by network (MV and NMV). Within each sub-plot we present the mean Frechet distance (± sem) between twin pairs (MZ, DZ) when responding with the Right or Left hand. Statistical analyses shows that response hand does not exert a significant effect on the Frechet distances across twin pairs. Therefore, further analyses of Frechet distance focused on effects of zygosity and network type, by collapsing the data across response hand.

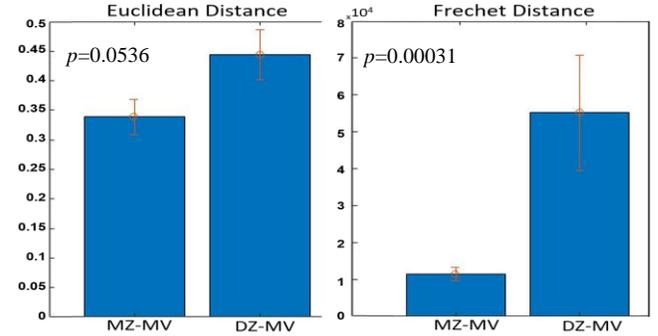

**Figure 2.** Comparison of Euclidean and Frechet distances considering the MV task-related subnetwork only.

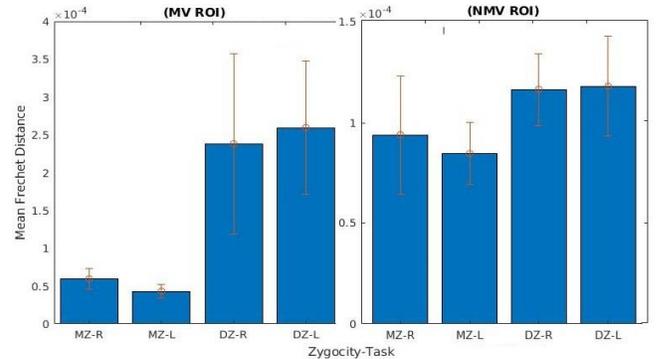

**Figure 3.** Mean Frechet distance of groups (MZ & DZ) for Right hand (R) & Left hand (L) tasks.

As seen in Figure 4 (with statistical analysis presented in Table 1), the mean Frechet distance between MZ is smaller than DZ for the MV but not for the NMV network. Effects using Euclidean distance were comparable for the MV network but, Euclidean distance was not sensitive to the effect of zygosity.

These findings suggest that, the MV (but not the NMV) sub-network is more similar in MZ compared to DZ. These results promote inter-related procedural and scientific inferences: First, Frechet distance is a geodesic metric that successfully identifies the effects of zygosity on brain network profiles in MZ and DZ twin pairs. Second, these results support the hypothesis that genetic similarity has an impact on brain network similarity, especially when the subjects perform the same task

### 4. DISCUSSION & CONCLUSION

In this paper, we have presented a computational framework to process the fMRI data and assess the similarity between brain networks. Specifically, the novel approach computes the difference between two graphs by finding the graph Laplacian of FC matrices (positive values only) and then applying Frechet distance. The advantage of this geodesic distance is that it accounts for the location and ordering of points in graphs. Also for this distance precludes the need to convert positive semi-definite matrices into positive definite by means of a regularization term (a process that can affect the results). This allowed us to discover scientifically relevant questions related to genetics, and its impact on brain network function.

**Table 1**. Statistical analysis results for comparing subnetworks (MV & NMV) of two groups (MZ & DZ).

| | Wilcoxon Ranksum Test | |
|---|---|---|
| **Pair** | **P-Value** (Euclidean Dist.) | **P-Value** (Frechet Dist.) |
| **DZ-MV vs MZ-MV** | 0.052643 | 0.000318 |
| **DZ-NMV vs MZ-NMV** | 0.630785 | 0.117926 |
| **DZ vs MZ** | 0.079542 | 0.000148 |

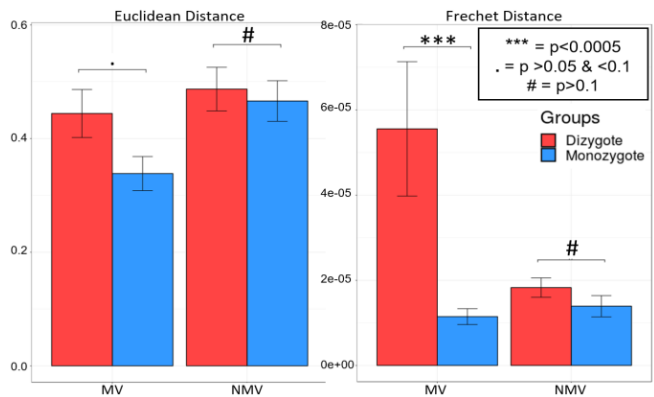

**Figure 4.** Connectivity similarity according to subnetworks (visuomotor on left and non-visuomotor on right) & Zygocity (Red=Dizygotic & Blue=Monozygotic).

The results of our study demonstrate how our analytic innovations reveal genetic influences on brain network profiles. In monozygotic twin pairs, the task-relevant visuomotor networks are more similar than they are in dizygotic twin pairs. On the other hand, there is no significant difference between these two groups when considering the task-irrelevant non-visuomotor network. These findings imply that zygosity modulates the connectivity of task-relevant networks, emphasizing a value of task-based fMRI.

**Acknowledgement**: The authors acknowledge Minh Ha Quang for the helpful discussions.